\documentclass{amsart}

\title{Is there a universal concept of mass in fundamental physics?}
\author{Robert Arnott Wilson}
\address{Queen Mary University of London}
\email{R.A.Wilson@qmul.ac.uk}

\date{First draft: 10th May 2022. This version: 19th June 2023.}

\begin{document}

\begin{abstract}
The concept of mass was introduced as a mathematical abstraction
and unifying principle in physics by Newton in the 17th century, and
calibrated on a Solar System scale by Cavendish at the end of the 18th century.
In the 19th century, this concept proved adequate to explain a vast range of
physical processes on all scales from the microscopic to the Solar System.
But in the 20th century, attempts to extend this range upwards to the
galactic scale, and downwards to subatomic particles, have led to
increasing difficulties. Modifications to the concept of mass by Einstein and Dirac
have not prevented these difficulties. In this paper, I ask the question,
can these difficulties be overcome by further modification of the
definitions, or is the concept of mass  an unavoidably local (Solar System scale)
rather than global concept?
\end{abstract}
\maketitle

\section{Introduction: inertia and gravity} 
It is hard to imagine any physics without the concept of mass. From Newton's laws of motion \cite{principia} onwards,
every physical theory is rooted in the concept of mass. Any suggestion that mass may not be a
universal property of matter is laughed at or ignored, since it is so obviously at odds with
centuries of experience in physics. Yet we must not forget that 
mass is first and foremost a
mathematical abstraction, designed by Newton to reduce physics to a series of mathematical
equations. The question of the physical `reality' of this `mass' is a philosophical question of
great difficulty.

Newton's laws can be analysed in terms of three separate concepts of mass, that is inertial mass
and active and passive gravitational mass. If we assume that inertial mass is the same
as passive gravitational mass, 
then universality of free fall follows from Newton's second law and his law of gravity.
Since experiment supports the universality of free fall to very high accuracy in a wide range of circumstances,
it is usual not to question this particular equivalence.
Nevertheless, universality of free fall does not imply a (global) equivalence of inertial and passive
gravitational mass, but only a local (and therefore approximate) equivalence, so that it is still possible to
cast doubt on this form of the Weak Equivalence Principle (WEP).

Similarly, Newton's law of gravity and his third law taken together imply that active and passive gravitational mass
are the same. In this case, however, experiment still struggles \cite{Gillies,newG}
to verify this assertion to better than about four-figure accuracy. Theoretically also,
it is not obvious that Newton's third law applies to gravitational action `at a distance', particularly
if we take the modern view that forces propagate at a finite rather than infinite speed.
Therefore the theoretical and experimental support for the equivalence of active and passive gravitational mass
is weak, and it is not unreasonable to question this version of the WEP.

Nevertheless, in the absence of definitive proof that active and passive gravitational mass are different,
most physicists are content to continue to treat them as equal, and most physical theories do not make
this distinction. Local experiments on a terrestrial or even Solar System scale have not produced the
kind of evidence that would force the issue. A few minor irritations like the anomalous perihelion precession of Mercury,
the Pioneer anomaly \cite{Pioneer1,Pioneer2} or the flyby anomaly \cite{flyby} have not been considered serious enough to overturn the
fundamental concept of mass.
In this paper, however, I shall distinguish two forms of the WEP, namely a weak form (WWEP), that is global equivalence of
inertial and passive gravitational mass, and a strong form (SWEP), that is the global equivalence of both of these with
active gravitational mass. I shall assume the WWEP, but not the SWEP, and aim to show that there is strong 
circumstantial 
evidence that the SWEP is in fact false.

Observations on a galactic scale \cite{Rubin} 
have demonstrated conclusively 
that the inertial mass of the
visible matter in the galaxy is much less than the active gravitational mass required to hold the stars in
their orbits. This may be because there is a lot of invisible (`dark') matter, or it may be because the (practical) equivalence
of inertial and active gravitational mass on a Solar System scale does not extend to the galactic scale.
Conventional wisdom supports the former assumption, although we still
lack direct evidence for
such dark matter \cite{DMreport}, and 
other theories have been put forward \cite{Milgrom1}.

The single most important experiment that supports the equivalence of inertial and active gravitational mass
is the Cavendish experiment \cite{Cavendish} of 1798. Modern experiments have improved the accuracy 
\cite{Gillies,newG,QLi} by two or three
significant figures, but that is all. Moreover, accurate Solar System astronomy can extend this equivalence
to the Solar System scale without much decrease in accuracy, since models of Solar System dynamics rely on the
active and passive gravitational masses of the planets being equal. Beyond that range, however,
tests of the SWEP are essentially impossible, 
since we cannot get a long way
from the Solar System to make direct tests. With increased accuracy, local tests on the Earth might 
eventually detect a clear signal of a difference between inertial and active gravitational mass,
but this has not happened yet. It is also possible that the equivalence principle (SWEP) actually does hold to high
accuracy on this scale, and such experiments will never find a discrepancy.

There is however another way to test whether the calibration of mass on a Solar System scale produces a universal
concept of mass, or only a local practical mass. If the latter, then we should detect this locality by observing strange
coincidences between mass ratios and Solar System dynamics. If the former, no such coincidences should exist.
This is not an experimental test or `proof' in the sense usually used in physics, but it is a plausibility argument 
based on the number (and precision) of suspicious numerical coincidences that can be found.
Such scientific investigation of coincidences and correlations should not, of course, be confused with magical notions of
numerology or arithmancy.
Experimental evidence to date 
indicates that we probably will not see any such correlations at the level of complete atoms, but we might see them at the level of
subatomic particles, for example in the masses of electron, proton and neutron.

\section{Inside the atom: a strange coincidence}
Clearly (?) the inertial masses measured in elementary particle experiments do not depend on Solar System dynamics. But if the
calibration of inertial and gravitational masses has left a historical trace of gravitational mass in a measurement of
inertial mass, then we should see it. Let us therefore ask the question:
 has the inertial mass ratio of (for example)
proton to electron retained an imprint of its calibration against gravitational mass in the 1960s and early 1970s,
before the universal adoption of quantum electrodynamics rendered this calibration obsolete? 

If it has, then we should see 
the dimensionless
constant 
\begin{align}
&&m(p)/m(e)\approx 1836.15&&
\end{align}
(see \cite{CODATA2018}) occurring somewhere obvious in Solar System dynamics.
 After a little thought, one might notice that
\begin{align}
&&1836.15 \approx 2 \times 365.24 /\sin(23^\circ 26' 33.7'')&&
\end{align}
in which 365.24 is the average number of days in a year, and $23^\circ 26' 33.7''$ was the angle of tilt of the Earth's axis
\cite{almanac}
in (approximately)
August 1957, June 1963 and March 1973.
Does this coincidence mean anything, and if so, what does it mean? 
We can either dismiss it as a meaningless coincidence, or take it as (circumstantial) evidence that the
Newtonian concept of mass, as calibrated by Cavendish, is irredeemably a local concept, that cannot
be meaningfully extended outside the Solar System.

It is worth making a rough estimate of the degree of `surprise' that this coincidence should elicit. 
First note that we could have chosen to use a sidereal year of 365.2564 days or an anomalistic year of 365.2596
days rather than a tropical year of 365.2422 days, which shifts the angle by about $4''$ and shifts the dates by a few years.
We could also have chosen to use a sidereal day of 23.9344696 hours, which shifts the angle by $4'$ and the dates by about
600 years.
If we now assume \emph{a priori} that the
electron mass could be anywhere between zero and the difference between the neutron and proton masses,
and that the number of days in a year is fixed,
then the surprise value here is roughly the surprise value of picking two random angles between $0^\circ$ and $90^\circ$,
and finding they differ by only a few seconds of arc. 
There is therefore roughly a $1$ in $10000$ chance that 
this is just a coincidence, without any underlying meaning.
Hence it is a strong signal, but not definitive, since physicists normally require a 1 in a million chance
before they will take a coincidence seriously. We therefore need to gather more evidence.

First, a word on accuracy. We are not talking here about a single experiment to measure a
single parameter, that can be dated to a particular time, but a large number of experiments conducted over 
a period of years to measure many interdependent parameters \cite{1969,1973}. Individual experiments can of course produce
highly accurate results, but the overall exercise is only as strong as its weakest link, which in this case is the direct measurement
of gravitational mass. At the time, the 1969 NIST recommendations  \cite{1969} quoted a standard uncertainty of 460 
parts per million (ppm) for the value of the Newtonian gravitational constant $G$, that was \emph{increased} to 730 ppm four years later  \cite{1973},
so that the consistency of the whole picture at that time cannot be
assumed to be any better than this. 
This level of uncertainty in the gravitational masses of proton and electron covers
a range equivalent to several centuries of variation in the angle of tilt of the Earth's axis. 

The fact that the experimental evidence that I have summarised here
locates the effective calibration to the correct period of two decades is therefore more than could have been 
reasonably expected.
Clearly the observed coincidence  does \emph{not} mean that the proton/electron
mass ratio changes as the angle of tilt of the Earth's axis changes: the inertial mass ratio is experimentally determined to much
higher accuracy than this, and does not change,
while the gravitational mass ratio cannot be directly measured at all. But it does mean that it is not possible to unify the theories of particle
physics and gravity on the basis of a single concept of mass. Any such unified theory \emph{must} separate at least two
distinct concepts of mass, as is done for example in \cite{ZYM}. It might be argued that Einstein's theory of General Relativity (GR)
already does this, by distinguishing between rest mass and total energy, but that is not enough, because we need to distinguish
two different kinds of rest mass. One of them changes as the rest frame of the laboratory changes due to the motion of the Earth,
while the other does not change in this way.

If the above conclusions are correct, then we should expect to see further traces of Solar System dynamics in other mass ratios. 
Again it must be emphasised that this does \emph{not} imply that the mass ratio changes as the Solar System dynamics change.
These traces must 
be interpreted instead as historical records of the calibration of various types of mass against each other, and therefore
represent choices made in the building of models, rather than anything fundamental about the underlying physics.
A detailed analysis of the historical record for the case of the proton/electron mass ratio is carried out in
\cite[Section 9]{icosa}, in which a correlation between the measured mass ratio and the angle of tilt is found, and provides
additional evidence that the hybrid practical mass in use before 1973 was effectively replaced 
in particle physics by a pure inertial mass after that date.

\section
{Inside the atomic nucleus: the plot thickens}
Let us first set reasonable bounds for which mass ratios we will accept as capable of providing meaningful evidence.
We surely cannot admit quarks, but only particles that have well-defined independent existence for long enough
that meaningful measurements can be made. Then we must include \emph{all} of the first-generation particles of spin $1/2$ and spin $0$, 
that is, besides the
proton and electron, the neutron, the electron neutrino and the pions. 
There are far too many second-generation mesons and baryons altogether, but if we want more particles, 
we can consider adding in the 
kaons, that is the `strange' mesons that are next in mass (a little more than half the mass of a proton) after the pions. 

This gives us a total of seven potentially independent mass ratios, 
of which four can be taken between charged and neutral versions of the same type of particle.
The other two mass ratios, between three very different types of particles (matter particles, pions and kaons),
have no obvious rationale and
will not be considered in this paper.
In each case we can put either the lighter or the heavier particle in the denominator, so can choose
a dimensionless constant either in the range $0$ to $1$, or in the range $1$ to infinity.
In order to estimate the unlikeliness of the coincidences, we need a prior distribution on this range, 
for example the uniform distribution on the range $0$ to $1$, but I shall make no attempt at rigorous analysis of probabilities in this paper.
There is little point in keeping much more than four figure accuracy here,
but for the sake of transparency I give five figures everywhere, and six where available \cite{CODATA2018,PDG}. 

The numerical values of the measured mass ratios are thus 
\begin{align}
&&m(p)/m(n) & \approx .998624&&\cr
&&m(\pi^0)/m(\pi^\pm) & \approx .967092&&\cr
&&m(\nu_e)/m(e)  & \approx .00000&&\cr
&&m(K^\pm)/m(K^0) & \approx .99209&&
\end{align}
For all of these ratios, or at least the non-zero ones, 
we seek simple functions of an angle $\theta$ or a frequency $f$. Since frequencies can be converted
naturally into angles by the standard formula $\theta=(360/f)^\circ$, there is no mathematical distinction between the two.
We may however want to keep a pair of frequencies, in order to distinguish between sidereal and synodic periods of time.
The kinds of functions we are looking for are very simple functions like
\begin{align}
&&\theta, 1-\theta, \sin\theta, \cos\theta, \sin^2\theta, \cos^2\theta&&
\end{align}
possibly with one or two normalisation constants such as $2$ or $\pi$. 

So far we have used the angle of tilt of the Earth's axis, where a nominal 
or historical value of $23.4427^\circ$
was used, compared to the current value of $23.4367^\circ$.
We have also used the frequency of the night and day oscillation over a complete year, $365.24$, which converts to an angle of
$.98565^\circ$. The next two available parameters of the same kind
are the inclination of the Moon's orbit, which has an average value of $5.145^\circ$, varying between
$4.99^\circ$ and $5.30^\circ$, and the night/day frequency over a complete month,
which is around $29.53$, and converts to an angle of $12.19^\circ$.

Our analysis of the electron/proton mass ratio shows that there is no real choice for the formula  in the proton/neutron case, which must be one of
\begin{align}
&&1-1/(2\times 365.24) & \approx .998631,&&\cr
&&1-1/(2\times 366.24) & \approx .998635.&&
\end{align}
Either of these gives us an accuracy of $10^{-5}$, which immediately increases the surprise value to a billion to one, 
although it must be emphasised that this is not a rigorous statistical analysis, and must be taken with a pinch of salt.
In particular, it assumes that the two surprises are statistically independent of each other, which may not be the case.

The other specific values we are looking for can be found in the following places:
\begin{align}
&&1-1/29.53 & \approx .966136&&\cr
&&1-1/(1+29.53) & \approx .967245&&\cr
&&\cos^2(5.145^\circ) & \approx  .99196&&\cr
&&\cos^2(5.30^\circ) & \approx .99147&&\cr
&&\cos^2(4.99^\circ) & \approx .99243&&
\end{align}
The second of these formulae gives the pion mass ratio to an accuracy of 150 ppm, and the third gives
the kaon mass ratio to 130 ppm.
The surprise value here is less easy to estimate, since I have not provided any \emph{a priori} reason for choosing one
formula rather than another.
Taken at face value, we get an overall $p$-value of around $10^{-17}$, however. 

I have not included any surprise value in the near-zero neutrino masses, but it could be claimed \emph{a priori} that 
the electron neutrino mass could be anything up to 
\begin{align}
&&m(n) - m(p) - m(e), &&
\end{align}
which is more than $1.5$ electron masses.
With this as the prior, rather than the original standard model assumption of zero mass, we would get a further factor of $10^{-5}$.
But I do not claim that a near-zero neutrino mass counts as a surprise, so exclude
this case.

\section{Conclusion: negative}
The value of this evidence can certainly be disputed, and it is not my purpose here to argue about exactly how likely or unlikely
such a set of coincidences really is. But if these mass ratios are really fundamental constants of nature,
then we should be very surprised to see even one such coincidence, let alone four. 
{\em Any two of them}, taken together, if they are independent, easily reach the one-in-a-million threshold for a new discovery.
Since these coincidences obviously
do not tell us anything about fundamental physics, they must tell us about our model-building instead.
What they appear to tell us is that basing our model-building 
for one-third of a millennium
on the 
assumption that inertial and gravitational mass are the same thing
may have been a mistake.
Could this be the underlying false assumption that 
the experts \cite{penrose,Smolin}
assert must exist? 

The assumption that there is a universal concept of `mass' that unites inertial forces and gravity was a major breakthrough
in the late 17th century, that underpinned a thorough mathematicisation of physical science, that reigned supreme for 200 years.
But cracks started to appear in the edifice in the early 20th century, when this assumption led to 
apparently insurmountable difficulties in uniting 
quantum mechanics 
with the theory of gravity, whether that be general relativity or
some other theory. The fundamental difficulty seems to be that Newton's law of gravity, in natural units, that is $G=c=1$,
where $G$ is the Newtonian gravitational constant and $c$ is the speed of light,
identifies mass with length, whereas quantum mechanics with $h=c=1$, where $h$ is Planck's constant, identifies mass with inverse length.
Almost a century later, these difficulties appear no nearer to a resolution.

There may be many reasons for this, but foremost amongst them is perhaps the fact 
that the universality assumption for mass no longer appears to be consistent with the experimental evidence.
The various superimposed rotations of our motion through the Solar System appear to have a dramatic effect on how we see
the universe, on all scales, as if `through a glass, darkly'.
 The impossibility of shielding any experiment from gravity
makes this hypothesis hard to refute.

The evidence I have presented in this paper suggests that the Standard Model of Particle Physics (SMPP) has been built
on top of the particular gravitational background that we experience on Earth, and that the numbers we put into the model
are (at least in some cases) taken from this background, rather than being universal constants. This does not mean that the
model is wrong, just that the mass coordinates that we use in the model are not the only possible ones. In other words, these mass
coordinates, however many (two or more) 
there may eventually turn out to be, can change in much the same way that spacetime coordinates can change in GR \cite{thooft,GR1,GR2,GRhistory}, 
without having any effect on the
underlying laws of fundamental physics.

\end{document}